\documentclass[10pt]{emulateapj}
\newcommand{\beq}{\begin{equation}}
\newcommand{\eeq}{\end{equation}}
\newcommand{\bea}{\begin{array}}
\newcommand{\eea}{\end{array}}

\shorttitle{Mean motion resonance formation in the planetary systems}
\shortauthors{Wang \& Ji}

\begin{document}

\title{Near 3:2 and 2:1 mean motion resonances formation in the systems observed by Kepler}

\author{Su Wang\altaffilmark{1} and Jianghui Ji\altaffilmark{1}}
\altaffiltext{1}{Key Laboratory of Planetary Sciences, Purple
Mountain Observatory, Chinese Academy of Sciences, Nanjing 210008,
China; wangsu@pmo.ac.cn, jijh@pmo.ac.cn.}

\begin{abstract}
The Kepler mission has released $\sim$ 4229 transiting planet
candidates. There are approximately 222 candidate systems with three
planets. Among them, the period ratios of planet pairs near 1.5 and
2.0 reveal that two peaks exist for which the proportions of the
candidate systems are $\sim$ 7.0\% and 18.0\%, respectively. In this
work, we study the formation of mean motion resonance (MMR) systems,
particularly for the planetary configurations near 3:2 and 2:1 MMRs,
and we concentrate on the interplay between the resonant
configuration and the combination of stellar accretion rate, stellar
magnetic field, speed of migration and additional planets. We
perform more than 1000 runs by assuming a system with a
solar-like star and three surrounding planets. From the statistical
results, we find that under the formation scenario, the proportions
near 1.5 and 2.0 can reach 14.5\% and 26.0\%, respectively. In
addition, $\dot M=0.1\times 10^{-8}~M_\odot~{\rm yr^{-1}}$ is
propitious toward the formation of 3:2 resonance, whereas $\dot
M=2\times 10^{-8}~M_\odot~{\rm yr^{-1}}$ contributes to the
formation of 2:1 resonance. The speed-reduction factor of type I
migration $f_1\geq 0.3$ facilitates 3:2 MMRs, whereas $f_1\geq 0.1$
facilitates 2:1 MMRs. If additional planets are present in orbits
within the innermost or beyond the outermost planet in a
three-planet system, 3:2:1 MMRs can be formed, but the original
systems trapped in 4:2:1 MMRs are not affected by the supposed
planets. In summary, we conclude that this formation scenario will
provide a likely explanation for Kepler candidates involved in 2:1
and 3:2 MMRs.

\end{abstract}


\section{Introduction}
\begin{figure*}
\begin{center}
  \epsscale{1.2}\plotone{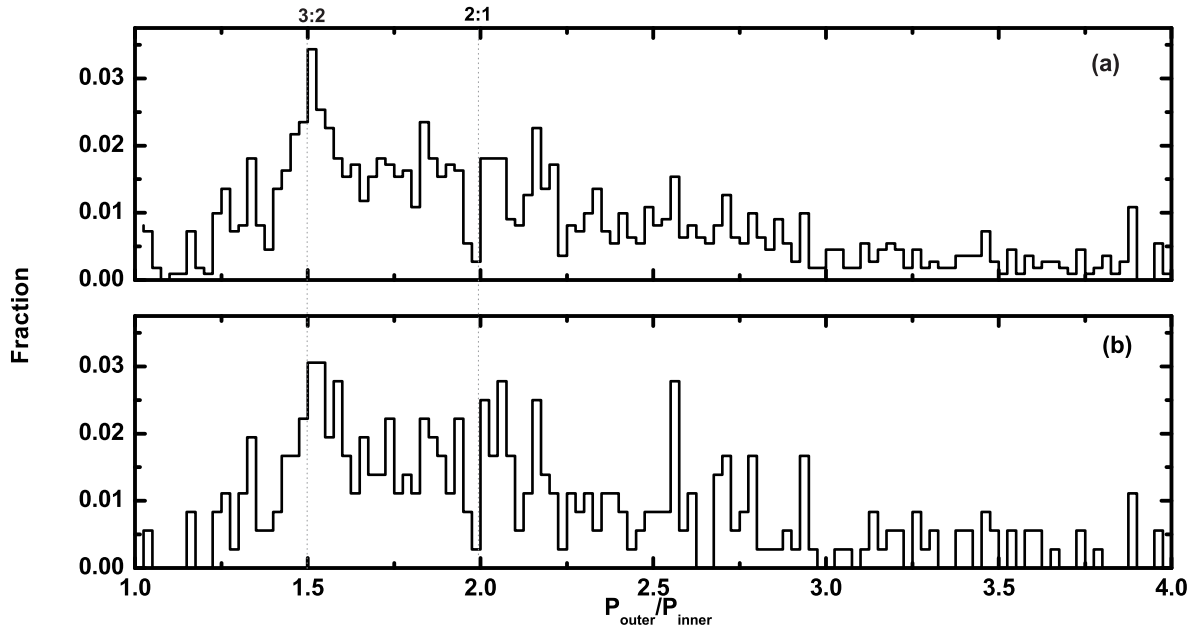}
 \caption{The statistical results of planet pairs in Kepler
 systems. Upper panel shows the period ratio of two planets for all
 planetary candidates, whereas the lower panel represents the values
 simply for three-planet systems. Two grey dotted lines are associated with exact 3:2
and 2:1 MMRs.
 \label{f1}}
 \end{center}
\end{figure*}

The Kepler mission has released data spanning over 16 months
\citep{Fab12,Bata13,Mazeh13}. The statistical results, obtained from
361 multiple-planet systems, show that there are two peaks for the
period ratios of two planets in a system: approximately 1.5 and 2.0.
This finding may provide evidence that there are a plenty of planet
pairs near 3:2 mean motion resonances (MMRs) (with the period ratio
of two planets in the range of [1.45, 1.54]) and 2:1 MMRs (with the
period ratio of two planets in the range of [1.83, 2.18])
\citep{Lissauer11}. As of July 2014, approximately 4229 planet
candidates have been reported in 2804 planetary systems, most of
which will be subsequently confirmed as authentic planets through
follow-up observations and by double checking the data. In this
population, there are 974 multiple planetary systems, including 652
two-planet systems, 222 three-planet systems and 75 four-planet
systems. As previously mentioned, most of the planetary candidates
in the multiple planetary systems are believed to be real planets
\citep{Lissauer12,Ciardi13,Qu13}. Based on current data, in Figure
\ref{f1}, we show the distribution of the period ratios of the
planet pairs for all of the planetary candidates (\textit{the upper
panel}) and for all of the three-planet systems (\textit{the lower
panel}). The upper panel shows that the proportion of the period
ratios for two planets near 1.5 ($\sim$ 10.5\%) and 2.0 ($\sim$
20.5\%) appear to be much larger than the proportion of other period
ratios. In contrast, the lower panel shows that two peaks are also
observed for the case of three-planet systems (see Fig.\ref{f1}) in
which the proportions of systems near 1.5 and 2.0 are approximately
7.0\% and 18.0\%, respectively. Furthermore, we note that two peaks
occur at period ratios greater than the exact values of 1.5 and 2.0.
Such a distribution may be related to planets near 4:2:1 MMRs
(so-called Laplacian resonances) or 3:2:1 MMRs.

These observations indicate that a great many systems are
approximately involved in MMR (Figure \ref{f1}). Therefore,
investigation into the formation scenarios of such resonant
configurations should inform planetary formation theory in several
respects. Currently, several formation scenarios are proposed to
explain the formation of systems near MMRs.

First, it is difficult for planet candidates observed by Kepler to
form in situ based on the Minimum Mass Solar Nebula (MMSN) model
\citep{hay81}. For all planetary candidates, the planetary radii are
in the range of [0.24, 154] $R_\oplus$ with an average value of 2.09
$R_\oplus$, whereas the radii of the planets in three-planet systems
range from 0.31 to 26 $R_\oplus$ with an average value of 1.92
$R_\oplus$. In addition, for three-planet systems, $\sim$ 75\% of
the planetary radii are in the range of [1, 3] $R_\oplus$ with
semi-major axes that are shorter than 1.14 AU. A recent
investigation of the internal structure of terrestrial and
Neptune-like planets indicates that a planet's mass radius follows
the relationship $R\propto M^{0.226-0.262}$
\citep{Val06,Val07,Marcus10,How13,lee13}. Thus, the planets have
average masses in the range of [12, 18] $M_\oplus$. Considering an
isolation mass $m_{\rm iso}$ in a core-accretion model
\citep{idalin04}, to yield a planet with the average mass and radius
of a three-planet system, the enhancement factor of the MMSN should
be greater than 16. Hence, it is difficult to explain the final
configuration of multiple-planet systems based on an in
situ-formation scenario.

The most classical theory for producing such a configuration is the
convergent migration scenario that occurs in the gaseous disk
\citep{GT80,lin96}. Given an appropriate orbital migration speed,
two planets can easily be captured in MMR \citep{Bryden00,MS01}.
Systems such as GJ 876, with two planets of a four-planet system
captured in a 2:1 MMR \citep{lee02,Ji02,Ji03,Zhou05,Zhang10}, and
KOI-152, which consists of three companions in a near-Laplacian
configuration \citep{Wang12}, are well explained via this migration
scenario, and the KBOs which are captured with Neptune in 3:2 or 2:1
MMRs are demonstrated by the migration of Neptune \citep{Mal95}.

An alternative scenario is proposed \citep{Pet13} to elucidate the
formation of near-MMR for systems with growing-mass planets. In this
case, the authors show that the 3:2 resonance is the strongest
first-order resonance for a planet with a mass of 20-100 $M_\oplus$.
However, the effects of dissipation orbital migration are not fully
taken into account in the authors' work, which may play a
significant role in the capture of resonant planets, especially for
a 2:1 MMR.

Recently, \citet{Ogihara13}, using numerical simulations,
investigated the formation scenario that triggered two planets to
capture a first-order MMR. In their work, the authors also
considered the combination of orbital migration and gas damping. For
two well-separated planets, the planets may undergo convergent
migration and may ultimately become trapped in MMR during their
evolution, whereas for closely spaced planet pairs, the planets may
be formed in situ. In their simulations, the authors explored simple
systems with two equal-mass planets. Nevertheless, the formation
scenario in a three-planet system would be quite different because
of the gravitational perturbation caused by the third planet.
Moreover, the planetary mass ratio acts as an alternative factor in
shaping the final configuration of the system. In-depth
investigation is required to understand planetary formation in
packed systems.

Based on our former study \citep{Wang12} and observational data, our
main aim in this work is to determine the formation scenario that
may lead to the various types of MMR that match statistical outcomes
by taking three-planet systems as an example. In our simulations, we
consider three planets whose masses are lower than 30 $M_\oplus$.
These low-mass planets are believed to form at a distant region
rather than at their present locations; thus, they will undergo type
I migration until they reach the inner region of the gaseous disk
and stop migrating. Subsequently, tidal interactions between the
planets and the central star will circularize their resultant
orbits. Such a formation process is mainly affected by three
important factors: the speed of the type I migration, the stellar
magnetic field, and the stellar accretion rate, which may shape the
final configuration of the investigated systems. Moreover, the
stellar magnetic field and stellar accretion rate affect the density
profile of the gaseous disk \citep{KL07,KL09,Wang12}, which also
plays an important role in determining each planet's final location.

In this work, we focus on the configuration formation of near-MMR,
particularly 3:2 and 2:1 MMRs, in three-planet systems as a function
of (1) the speed of the type I migration based on the value obtained
by linear analysis, (2) the stellar accretion rate from early to
later stages, (3) the stellar magnetic field, and (4) the potential
survival of additional planets in the systems that may break up the
MMRs. In Section 2, we summarize our models, including the disk
model and the migration scenario that planets may experience.
Section 3 presents the numerical results obtained for two different
cases of planet configuration. Section 4 presents our discussion and
conclusions.

\section{Models}
\subsection{Disk Model}
The surface density of a gas disk at a stellar distance $a$ is
described as \citep{Pri}

\begin{equation}
\Sigma_{\rm g}=\frac{\dot M}{3\pi \alpha (a) c_s h}{\rm
exp}\left(\frac{-t}{\tau_{\rm dep}}\right)\eta, \label{dens}
\end{equation}
where $t$ is time, the timescale $\tau_{\rm dep}$ is estimated to be
approximately several million years \citep{hai01} and $c_s$ and $h$
are the speeds of sound at the midplane and isothermal density scale
height, respectively. Herein, $\dot M$ is the stellar accretion
rate, which can be evaluated as \citep{natta,Vor09}

\begin{equation}
\dot M\simeq 2.5\times
10^{-8}\left(\frac{M_*}{M_\odot}\right)^{1.3\pm 0.3}M_\odot~ {\rm
yr}^{-1}. \label{mdot}
\end{equation}
For a star with a solar mass, the stellar accretion rate is
$\sim2.5\times 10^{-8}$$M_\odot~ \rm {yr^{-1}}$. The average value
of this rate decreases as the star and the disk evolve. Thus,
herein, we suppose that the stellar accretion rate varies from
$0.1\times 10^{-8}$ to $2.5\times 10^{-8}$ $M_\odot~ \rm {yr^{-1}}$
\textbf{representing different stage of the star evolution} for a
solar-like system. In addition, the efficiency factors of angular
momentum transport $\alpha$ and $\eta$ are defined as

\begin{equation}
\alpha_{\rm eff} (a)=\frac{\alpha_{\rm dead}-\alpha_{\rm mri}}{2}
\left[{\rm erf}\left(\frac{a-a_{\rm crit}}{0.1 a_{\rm
crit}}\right)+1\right]+\alpha_{\rm mri},
 \label{alpeff}
\end{equation}
\begin {equation}
\eta=0.5\left[{\rm erf}\left(\frac{a-a_{\rm mstr}}{0.1a_{\rm
mstr}}\right)+1\right],
\end{equation}
where $a_{\rm crit}$ and $a_{\rm mstr}$ are the locations of the
boundary of magneto-rotational instability (MRI) and the truncation
of the magnetic field, respectively. Specifically, the two
parameters are modeled as \citep{KL09,konigl}

\begin{eqnarray}
a_{\rm crit}=0.16 ~{\rm
AU}~\left(\frac{\dot{M}}{10^{-8}M_{\odot}~{\rm
yr}^{-1}}\right)^{4/9} \left(\frac{M_{*}}{M_{\odot}}\right)^{1/3}
\nonumber\\
\times\left(\frac{\alpha_{\rm
mri}}{0.02}\right)^{-1/5}\left(\frac{\kappa_D}{1{\rm
cm^2g^{-1}}}\right),~~~~
 \label{acrit}
\end{eqnarray}

and

\begin{eqnarray}
a_{\rm mstr}=(1.06\times 10^{-2} ~{\rm AU}) \beta'
\left(\frac{R_*}{R_\odot}\right)^{12/7}\left(\frac{B_*}{1000{\rm
G}}\right)^{4/7}
\nonumber\\
\times\left(\frac{M_*}{M_\odot}\right)^{-1/7}\left(\frac{\dot{M}}{10^{-7}M_\odot~
{\rm yr^{-1}}}\right)^{-2/7},~~ \label{mstr}
\end{eqnarray}
where $\kappa_D$ and $B_*$ are the grain opacity and stellar
magnetic field, respectively. With respect to spherical accretion,
$\beta'=1$ represents a typical Alfv\'{e}n radius. The variables
$\alpha_{\rm dead}=0.001$ and $\alpha_{\rm mri}=0.01$ denote the
value of $\alpha$ in the dead zone and in the active zone at the
midplane of the disk, respectively.

Based on the above-described equations, we can calculate the gas
density $\Sigma_g\propto r^{-1}$. The stellar accretion rate and
stellar magnetic field are believed to be two important factors that
affect the profile of the gas density \citep{Wang12}. Therefore, in
our model, we consider these parameters in the investigation of
planet formation and test our simulations by varying them.

\subsection{Eccentricity damping and planetary migration }
For a planetary embryo embedded in a gas disk, mutual interactions
will result in the eccentricity damping of the embryo on a timescale
$\tau_{\rm damp}$ represented as \citep{Cre06}

\begin{eqnarray}
\tau_{\rm damp}=\left(\frac{e}{\dot
e}\right)=\frac{Q_e}{0.78}\left(\frac{M_*}{m}\right)
\left(\frac{M_*}{a^2\Sigma_g}\right)\left(\frac{h}{r}\right)^4\Omega^{-1}
\nonumber\\
\times\left[1+\frac{1}{4}\left(e\frac{r}{h}\right)^3\right] {\rm
yr},~~~~~~~~~~~~~~~~~ \label{damp}
\end{eqnarray}
where $Q_e=0.1$ is a normalized factor fitted to hydrodynamical
simulation results and $h,~r,~\Omega$, and $e$ are the disk scale
height, the distance from the central star, the Kepler angular
velocity, and the eccentricity of the embryo, respectively.

Additionally, the angular momentum exchange between embedded planets
and the gaseous disk will trigger orbital migration of the planets.
According to the core-accretion scenario \citep{idalin04}, type I
and type II migration are proposed to explain the formation of
close-in super Earths or hot Jupiters.

If a planet embryo occupies a low mass ($\leq$ 30 $M_\oplus$), the
angular momentum exchange between the embryo and gaseous disk can be
analyzed using a linear model, and the net loss from the embryo will
eventually lead to an inward migration \citep{GT79,Ward97,Tan02}.
The timescale for type I migration can be expressed as

\begin{eqnarray}
\tau_{\rm migI}=\frac{a}{|\dot{a}|}=\frac{1}{f_1(2.7+1.1\beta )}
\left(\frac{M_*}{m}\right)\left(\frac{M_*}{\Sigma_ga^2}\right)
\nonumber\\
\times\left(\frac{h}{a}\right)^2
\left[\frac{1+(\frac{er}{1.3h})^5}{1-(\frac{er}{1.1h})^4}\right]\Omega^{-1}\rm{yr},~~~~~~
\label{tauI}
\end{eqnarray}
where $e,~r,~h$ and $\Omega$ bear the same definition as in equation
(\ref{damp}). The variable $f_1$ is the reduction factor. Using the
gas-density profile given by equation (\ref{dens}), we have $\beta
=-d\ln \Sigma_g/d\ln a=1$.

\begin{table*}
\centering \caption{The initial parameters of the groups in Case 1.
 \vspace{0.5cm}
 \label{tb1}}
\begin{tabular*}{18cm}{@{\extracolsep{\fill}}lllll}
\tableline
& Group 1 & Group 2& Group 3& Group 4\\
\tableline
Mass of planets ($M_\oplus$)&(5, 10, 15)& (5, 5, 5), (5, 5, 10) & (2-30, 0.2-15, 0.5-22.5) & (5, 10, 15)\\
$\dot M$ ($\times 10^{-8}$$M_\odot$$\rm{yr^{-1}}$)&0.1-2.5  &0.1-2.5  & 0.1-2.5  & 0.1, 0.5\\
$B_*$ (KG)&0.5, 1, 1.5, 2, 2.5  &0.5, 1, 1.5, 2, 2.5  & 0.5, 1, 1.5, 2, 2.5 & 0.5, 1, 1.5, 2, 2.5\\
$f_1$&0.01, 0.03, 0.1, 0.3, 1&0.01, 0.03, 0.1, 0.3, 1 & 0.01, 0.03,
0.1, 0.3, 1 &0.01, 0.03, 0.1, 0.3, 1\\
 \tableline
\end{tabular*}
\end{table*}

\begin{table*}
\centering \caption{The initial parameters of the cases in Case 2.
 \label{tb2}}
\begin{tabular*}{10cm}{@{\extracolsep{\fill}}lll}
\tableline
 &A1 & A2\\
\tableline
Mass of planets inner ($M_\oplus$)  &2, 5  & 2, 5 \\
Mass of planets outer ($M_\oplus$)  &5, 20 & 5, 20\\
$\dot M$ ($\times 10^{-8}$$M_\odot$$\rm{yr^{-1}}$)&0.1, 0.5, 2 &0.1, 0.5, 2\\
$B_*$ (KG)&0.5, 1 &0.5, 1\\
$f_1$&0.03, 0.1&0.3, 1\\
\tableline
\end{tabular*}
\end{table*}
In this work, we consider the gravitational interaction between each
body in the system, type I migration (for planets with masses less
than 30 $M_\oplus$), and the eccentricity damping of the planets.
The total acceleration of the planets with mass $m_i$ is expressed
as

\begin{eqnarray}
\frac{d}{dt}\textbf{V}_i =
 -\frac{G(M_*+m_i
)}{{r_i}^2}\left(\frac{\textbf{r}_i}{r_i}\right) +\sum _{j\neq i}^N
Gm_j \left[\frac{(\textbf{r}_j-\textbf{r}_i
)}{|\textbf{r}_j-\textbf{r}_i|^3}- \frac{\textbf{r}_j}{r_j^3}\right]
\nonumber\\
+\textbf{F}_{\rm damp}+\textbf{F}_{\rm migI},~~~ \label{eqf}
\end{eqnarray}
where

\begin {eqnarray}
\begin{array}{lll}
\textbf{F}_{\rm damp} = -2\frac{\displaystyle (\textbf{V}_i \cdot
\textbf{r}_i)\textbf{r}_i}{\displaystyle r_i^2\tau_{\rm damp}},
\\
\cr\noalign{\vskip 0.5 mm} \textbf{F}_{\rm
migI}=-\frac{\displaystyle \textbf{V}_i}{\displaystyle 2\tau_{\rm
migI}},
\end{array}
\end{eqnarray}
where $\textbf{r}_i$ and $\textbf{V}_i$ represent the position and
velocity vectors of planet $m_i$, respectively, and all of the
vectors are expressed in stellar-centric coordinates.

To explore the dynamical evolution of the planets, we integrate
equations (\ref{eqf}) using the Hermit scheme \citep{Aarseth}, which
is a time-symmetric integrator.

For the numerical setup, we assume that the system contains a
solar-mass central star with a surrounding gaseous disk. Initially,
all of the planets should be in coplanar and near-circular orbits.
The argument of the pericenter, the mean anomaly, and the longitude
of ascending node are randomly generated to be between $0^{0}$ to
$360^{0}$. To examine the role of the stellar accretion rate,
stellar magnetic field, the speed of the type I migration, and
additional planets, we consider two cases with different initial
parameters. For Case 1, we mainly investigate the configuration
formation for three-planet systems with the various parameters we
have chosen. For Case 2, we examine how additional planets affect
the evolution of the system. In the following section, we will
briefly introduce the main results obtained from the numerical
simulations.

\begin{figure*}
\begin{center}
  \epsscale{1.2}\plotone{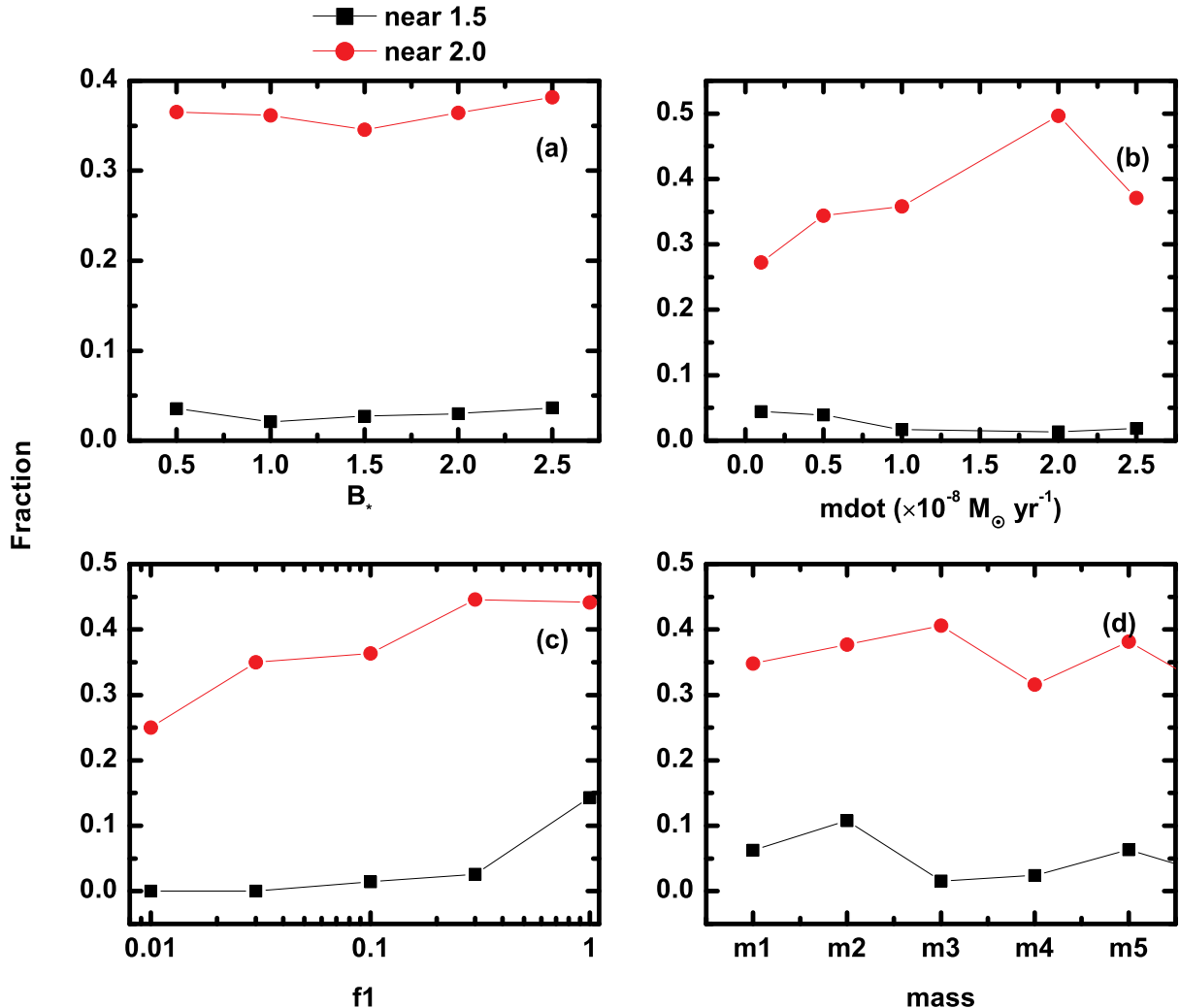}
 \caption{The statistic results of all runs in G1-G3.
 Panel (a) shows the distribution of period ratio with different
 star magnetic field fixed and other parameters are free. The black square represents the planet pairs near 3:2 MMR and the red
 circle means the statistics of planet pairs near 2:1 MMR.
 Panel (b) shows the results with different star mass accretion rate. Panel (c)
 displays the results of different speed of type I migration. Panel (d) represents the
 fraction of the period ratio varies with the masses of the three planets.
 \label{f2}}
 \end{center}
\end{figure*}

\section {Numerical Simulations and Results}
\subsection{Case 1}

We mainly consider MMR formation in three-planet systems. In this
case, we perform a total of 1020 runs. Furthermore, we choose
three planets with variable masses for G1-G3, and we also change the
initial locations for the planets for G4. Herein, we summarize each
of the initial systems for G1-G4 that are adopted in the numerical
simulations.

G1: In this group, given the isolation mass, the planet mass $m$ is
proportional to $a^{3/4}$. Therefore, the masses of the three
planets are chosen to be 5, 10, and 15 $M_\oplus$ (m1 labeled in
panel (d) of Figure \ref{f2}). For the first subgroup, the stellar
accretion rate ranges from $0.1\times 10^{-8}$ to $2.5\times 10^{-8}$
$M_\odot~\rm{yr^{-1}}$, and $f_1$ varies from 0.01 to 1, whereas
$B_*$ always remains 0.5 KG in this subgroup. For the second
subgroup, we fix the stellar accretion rate at $0.1\times 10^{-8}$
$M_\odot~\rm{yr^{-1}}$, whereas the stellar magnetic field ranges
from 0.5 to 2.5 KG, and $f_1$ varies from 0.01 to 1. The third
subgroup is quite similar to the second subgroup except that the
stellar rate remains $0.5\times 10^{-8}$ $M_\odot~\rm{yr^{-1}}$. In
summary, we carry out a total of 90 runs for G1 (including
three subgroups).

G2: For this group, three subgroups are also considered in the
simulations. The initial parameters adopted are similar to the
parameters adopted for G1 except for the planetary masses. The
masses of the three planets are divided into two cases: (1) the
three planets have an equal mass of 5 $M_\oplus$ (m2 labeled in
panel (d) of Figure \ref{f2}) or (2) the inner two planets both bear
a mass of 5 $M_\oplus$ but the mass of the outer companion is 10
$M_\oplus$ (m3 labeled in panel (d) of Figure \ref{f2}).
Additionally, the stellar accretion rate in subgroups 1, 2 and 3 is
assumed to be $0.1\times 10^{-8}$, $0.5\times 10^{-8}$, $1\times
10^{-8}$, $2\times 10^{-8}$, and $2.5\times 10^{-8}$
$M_\odot~\rm{yr^{-1}}$, respectively. Herein, we perform 130
runs for G2.

G3: The statistical results (see Figure \ref{f1}) show that most of
the Kepler planets are in near-MMRs, particularly 3:2 and 2:1 MMRs.
For three-planet systems, we classify the ratios of the planetary
masses into four types: 1:1:1.5, 2:1:7.5, 12:1:0.2 and 2:0.2:4. For
further investigation, we then choose six subgroups for three
planets depending on the variable masses such that the combinations
of the planetary masses are as follows: [5, 5, 7.5] $M_\oplus$
(subgroup m4), [2, 1, 7.5] $M_\oplus$ (subgroup m5), [30, 2.5, 0.5]
$M_\oplus$ (subgroup m6), [15, 15, 22.5] $M_\oplus$ (subgroup m7),
[6, 3, 22.5] $M_\oplus$ (subgroup m8) and [2, 0.2, 4] $M_\oplus$
(subgroup m9). Herein, the stellar accretion rates are assumed to be
$0.1\times 10^{-8}$, $0.5\times 10^{-8}$, $1\times 10^{-8}$, $2\times
10^{-8}$, and $2.5\times 10^{-8}$ $M_\odot~\rm{yr^{-1}}$,
respectively, whereas other the initial parameters are similar to
the parameters assumed for G1, except for the planetary masses.
Therefore, we carry out 750 runs for G3.

G4: To determine the likelihood that there is a high proportion of
planet pairs near 3:2 MMR, we reset the initial locations of the
three planets, which differ greatly from the locations considered in
G1-G3. We also choose five subgroups for the stellar accretion rate
(as performed for G1-G3), and the stellar magnetic field is set at
0.5, 1, 1.5, 2, and 2.5 G. The masses of the three planets are
exactly equal to the masses in G1. For this group, 50 runs are
carried out in the simulations.

For each of the above mentioned runs, we performed the
simulation over a timescale of 5 Myr. In the following section, we
will concisely summarize the most significant results obtained from
our numerical simulations.

\begin{figure*}
\begin{center}
  \epsscale{1.1}\plotone{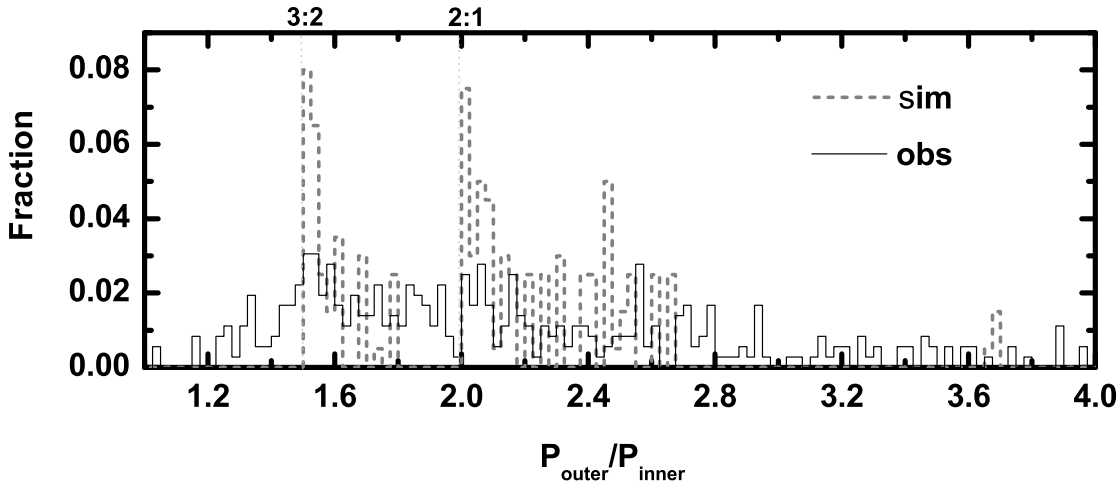}
 \caption{The statistic results of observation of three-planet systems and simulation results from 50 runs in G1 and
 50 runs in G4. The black solid line means the observation results and the
 grey dot represents the simulation results in G1 and G4.
 \label{f3}}
 \end{center}
\end{figure*}

\subsubsection{Statistic outcomes for G1-G3}

In groups G1-G3, the three planets possess orbital periods of 100,
250, and 600 days for $f_1$ in the range of [0.01, 0.03], and 140,
500, and 1450 days for $f_1 \geq 0.1$ initially.

From the statistical results obtained for G1-G3, we observe that the
proportion of planet pairs near a 2:1 MMR is approximately 36.4\%,
with the period ratio ranging from 1.83 to 2.18. However, the
proportion of planet pairs near a 3:2 MMR is approximately 3.0\%,
with the period ratios in the range of [1.45, 1.54]. Figure \ref{f2}
shows the statistical results obtained for period ratios near 1.5
and 2.0, and panels (a), (b), (c) and (d) present the results
obtained by considering variations in the stellar magnetic field,
stellar mass accretion rate, the speed of the type I migration, and
the planet mass. Based on Figure \ref{f2}, we may conclude that it
is difficult for two planets to be trapped near a 3:2 MMR under such
a formation scenario. However, two planets can be easily captured in
a 2:1 MMR. Hence, it is inevitable for three planets to be trapped
in 4:2:1 MMR through the aforementioned formation scenario based on
our model.

Panel (a) of Figure \ref{f2} displays the statistics for periods
near 1.5 and 2.0 with different stellar magnetic field. The red
dotted line represents the results obtained near a period ratio of
2.0, whereas the line composed of black squares represents the
results obtained near 1.5. Based on the results presented in panel
(a), we can conclude that the proportion corresponding to
approximately 2.0 is largest for $B_*=$ 2.5, whereas for $B_*=$ 0.5
and 2.5 the proportion near 1.5 is nearly the same, which is the
highest value obtained among the five scenarios. The proportion
changes slightly with an increase in the magnetic field, which
enables us to conclude that the stellar magnetic field plays a less
significant role in the formation of 2:1 or 3:2 MMRs over the
evolution of the system.

Panel (b) shows that, for the formation of 2:1 MMRs, the $\dot
M=2\times 10^{-8}~M_\odot~\rm{yr^{-1}}$ values are larger than the
values observed in other cases. Thus, $\dot M=2\times
10^{-8}~M_\odot~\rm{yr^{-1}}$ is the most favorable value for
producing 2:1 MMRs. To yield 3:2 MMRs, this proportion becomes
greatest when $\dot M=0.1\times 10^{-8}~M_\odot~\rm{yr^{-1}}$, as
shown in Figure \ref{f2}. Therefore, 2:1 MMRs can form easily during
the early stages of star formation with a high mass accretion rate,
whereas 3:2 MMRs can form easily during the late stages of star
formation. In cases in which the stellar accretion rate remains
unaltered, we also notice that several planet pairs are trapped near
higher-order MMRs of approximately 5:3, 5:2 and 3:1. Furthermore,
$\dot M=0.1\times 10^{-8}~M_\odot~\rm{yr^{-1}}$ is propitious for
the formation of planet pairs near 5:2 and 3:1 MMRs.

Panel (c) of Figure \ref{f2} shows the proportions of planet pairs
with different migration speeds. Based on the results presented in
Figure \ref{f2}, we can determine that the proportion of planet
pairs near a 2:1 MMR is greater than 35\% for $f_1 =$ 0.1, 0.3, and
1, and we also find that the proportion of planet pairs near a 3:2
MMR is greater than 2.5\% for $f_1 \geq$ 0.3, and greater than 10\%
for $f_1 =$ 1. Therefore, $f_1 \geq 0.1$ appears to favor planetary
configurations near-2:1 MMRs, which is consistent with the work of
Ida \& Lin (2008), Wang \& Zhou (2011) and Wang et al. (2012).
However, $f_1 \geq 0.3$ is propitious for the formation of near-3:2
MMRs. We also note that $f_1=0.03$ facilitates the capture of two
planets in near-3:1 MMRs.

\begin{figure*}
\begin{center}
  \epsscale{1}\plotone{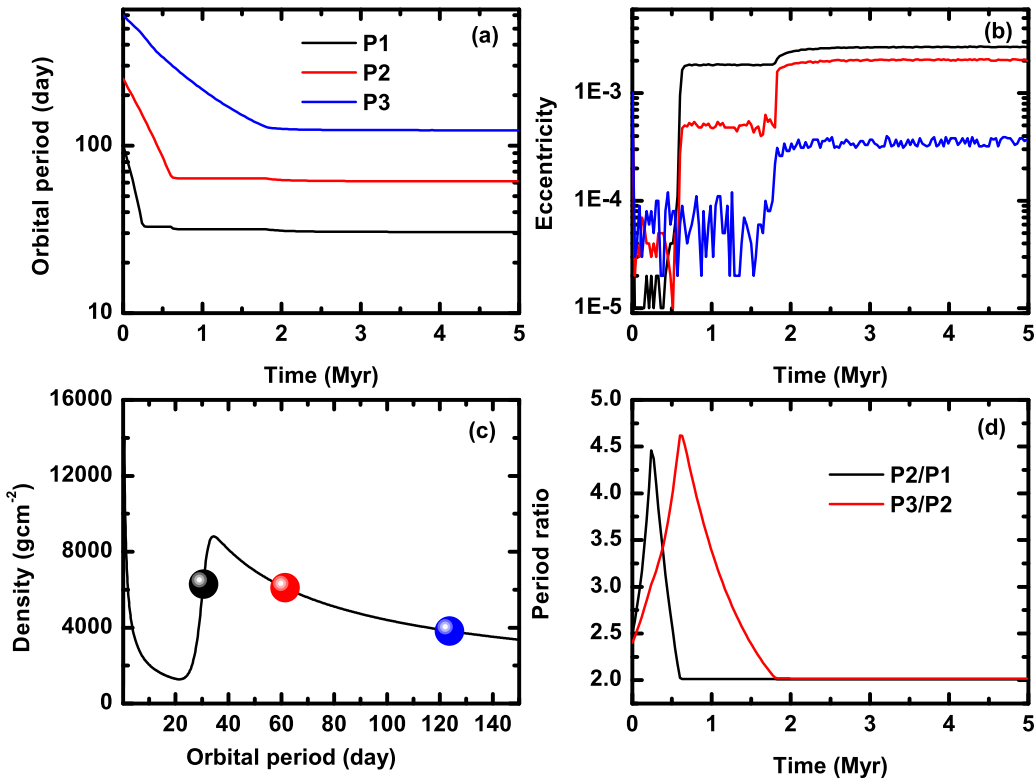}
 \caption{Results for R1. Panels (a), (b) and (d) show the
 evolution of orbital period, eccentricity, and period ratio of
 planets pair, respectively. Panel (c) shows the final configuration
 of three planets. P1, P2, and P3 represent the innermost, the intermediate and
 the outermost planet, respectively. The black solid curve is the
 density profile of the gas disk initially.
 \label{f4}}
 \end{center}
\end{figure*}

From panel (d) of Figure \ref{f2}, we observe that the proportion of
planet pairs forming near-2:1 MMRs is greater than 30.0\% for the
five scenarios, where subgroup m3 with the inner two planets being
of equal mass bears the largest proportion. For planet pairs forming
near-3:2 MMRs, the situation in which there are three equal-mass
planets, demonstrates the largest proportion. These results indicate
that two planets are easily involved in or are near a first-order
MMR if they have equal masses.

\subsubsection{Statistical outcomes for G4}

In G1-G3, the period ratios of the three planets are initially set
to be greater than 2.0. From the results given in Section 3.1.1, we
note that the fraction of planet pairs with period ratio near 1.5
from the simulations is lower than that from the observation.
Therefore, in G4, we set the systems with a compact structure by
varying the initial orbits of the three planets with the period
ratio, which ranges between 1.5 and 2.0, at 140, 270, and 530 days
initially. Thus, we perform 50 runs for G4 with those parameters in
Table \ref{tb1}. In these simulations, the planetary masses are
adopted to be 5, 10 and 15 $M_\oplus$, which are identical to those
in G1.

Upon the completion of the simulations, we collect the data obtained
from G4 (50 simulations with the initial parameters given in Table
\ref{tb1}) and \textbf{G1} (50 simulations with identically initial
parameters of the stellar magnetic field, the stellar accretion
rate, and the reduce factor of type I migration as in G4, but with
various initial locations), and the simulation results of the 100
simulations are plotted to compare them with the observations.

The statistical results are shown in Figure \ref{f3}. The black
solid line in this figure denotes the distribution profile of the
Kepler observations, whereas the grey dot represents the results of
our simulations. In Figure \ref{f3}, we note that there exist two
peaks near 1.5 and 2.0 from the numerical simulations that are
consistent with the observations. The proportion of planet pairs
near 1.5 and 2.0 is approximately 14.5\% and 26.0\%, respectively.
Compared with the results obtained for G1-G3 (see Figure \ref{f2}),
we find that the proportion near 1.5 increases, whereas the
proportion near 2.0 decreases. Furthermore, we notice that a lower
proportion peak emerges near 2.5 (5:2 MMR), but the planet pair is
absent at 3:1 MMR due to the limited number of simulations
performed. Moreover, we perform a KS test on the distributions, we
find that the p-value $p=0.06$ for $\dot M = 0.1 \times 10^{-8}$
$M_\odot~\rm yr^{-1}$ when the simulations compare with the
observation data, implying that they roughly have similar
distribution. Consequently, we conclude that the formation scenario
proposed in this work may reproduce a subset of the planet pair
distributions.

\subsubsection{R1: Two pairs of planets both in 2:1 MMR}

In this run, the initial parameters (see Table \ref{tb3}) are
assumed to be as follows: the masses of three planets are
identically set to be 5 $M_\oplus$; the stellar accretion rate is
adopted to be $1 \times 10^{-8}$ $M_\odot~\rm{yr^{-1}}$,which
indicates a middle stage for the star formation; the stellar
magnetic field is 0.5 KG; and the reduction factor $f_1$ is 0.03,
implying a slow speed of type I migration. Based on the simulation
results, we find that the three planets are trapped in 4:2:1 MMRs in
9.5\% of all simulations.

Figure \ref{f4} shows the typical evolution leading to the formation
of a Laplacian configuration. Panels (a) and (b) show the dynamical
behavior of the semi-major axes and eccentricities over a timescale
of 5 Myr. After the planets are captured in the resonances, the
semi-major axes remain nearly unchanged while the eccentricities
approach zero; here, the black, red and blue lines represent the
innermost, intermediate and outermost planets, respectively. In
addition, the trio's final configuration is clearly shown in panel
(c), which is indicative of the relationship between the final
orbits (denoted by the orbital periods) and the gas density profile.
Moreover, panel (d) shows the variations in the period ratio of two
planet pairs. Furthermore, in Figure \ref{f4}, we note that the
innermost planet reaches the maximum gas density very quickly at
$\sim$ 0.2 Myr, and the inner two planets are locked in a 2:1 MMR at
$\sim$ 0.6 Myr at the time the second planet arrives, but the outer
two planets are in a 2:1 MMR at $\sim$ 1.8 Myr. Again, based on
panel (c), we observe that the inner two planets cease migrating in
the regime in which the gas density nearly reaches its maximum, and
the outer two companions also stop moving inward, following the
behavior of the inner two. Ultimately, the three planets stop
migrating at orbital periods of 30.5, 61.3 and 123.6 days, from
innermost to outermost. The planets' eccentricities are damped due
to the high density of the gas disk. Such a configuration is
reminiscent of the KOI-1426 system, which also consists of three
planets with orbital periods of 38.9, 74.9 and 150.0 days, from
innermost to outermost, indicating that the system may be captured
into Laplacian resonance during dynamical evolution.

\begin{figure*}
\begin{center}
  \epsscale{1}\plotone{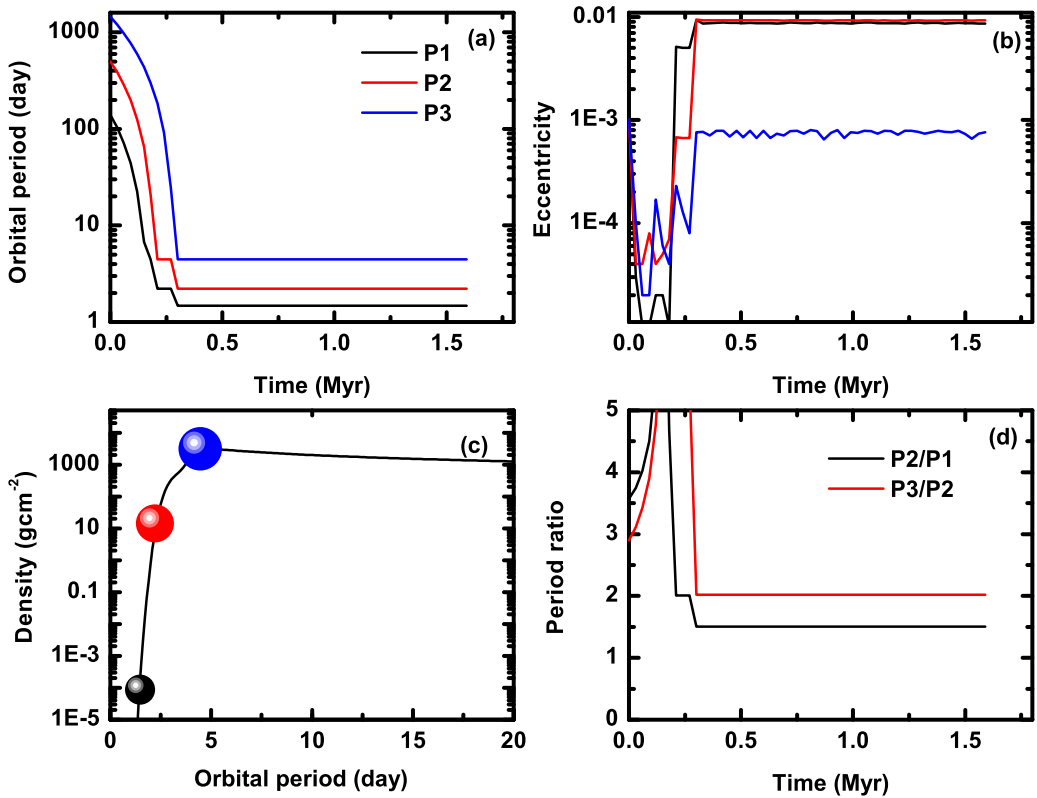}
 \caption{Results for  R2. Panels (a), (b), and (d) show the
 evolution of orbital period, eccentricity, and period ratio of
 the planet pair, respectively. Panel (c) shows the final configuration
 of three planets. P1, P2, and P3 denotes the innermost, the intermediate and
 the outermost planet, respectively. The black solid curve is the
 density profile of the gas disk initially.
 \label{f5}}
 \end{center}
\end{figure*}

\begin{table*}
\centering \caption{The initial parameters of R1, R2, R3 and R4.
 \label{tb3}}
\begin{tabular*}{14cm}{@{\extracolsep{\fill}}lllll}
\tableline
 &R1 & R2& R3& R4\\
\tableline
Mass of planets ($M_\oplus$)  &5, 5, 5 &5, 10, 15 &5, 5, 5 & 5, 5, 7.5\\
$\dot M$ ($\times 10^{-8}$$M_\odot$$\rm{yr^{-1}}$)&1&0.1&1 &0.1\\
$B_*$ (KG)&0.5&1&0.5&1.5\\
$f_1$&0.03 &1&0.3&1\\
\tableline
\end{tabular*}
\end{table*}

\subsubsection{R2: The inner pair is in a 3:2 MMR, whereas the outer pair is in a 2:1 MMR}
In R2, the masses of the three planets are 5, 10 and 15 $M_\oplus$,
the stellar accretion rate is  $0.1 \times 10^{-8}$
$M_\odot~\rm{yr^{-1}}$, and the stellar magnetic field is 1 KG. The
speed of the type I migration is equal to unity, which is consistent
with linear theory \citep{GT79,Ward97,Tan02}.

Based on the statistics obtained, we find that approximately 2.2\%
of the total simulations are associated with the final configurations
in which the inner two planets are in a 3:2 MMR, whereas the outer
two are trapped in a 2:1 MMR. Figure \ref{f5} shows the outcomes for
a typical case. Similar to Figure \ref{f4}, panels (a) and (d) show
that the inner two planets are locked in a 2:1 MMR at $\sim$ 0.2 Myr
initially. Subsequently, when the third planet P3 approaches P2, the
2:1 MMR is disrupted, and the inner pair is then trapped in a 3:2
MMR at $\sim$ 0.3 Myr, while the outer two companions enter a 2:1
MMR. Furthermore, panel (c) shows that the three planets can travel
across the region where the density reaches its maximum. In
addition, the outermost planet simply halts its migration in the
scenario in which the density is maximal. Moreover, the results show
that the eccentricities of the inner pair can be excited to 0.01
when they enter a 2:1 MMR, and the outermost planet's eccentricity
is slightly increased to 0.001 in the case of the arrival at
resonance, which is different from the case of R1. At the end of the
simulation, the three planets hold orbital periods of 1.5, 2.2 and
4.5 days. We also observe a comparable analog among the Kepler
candidates KOI-584, with orbital periods of 6.5, 9.9 and 21.2 days,
from innermost to outermost, by considering the orbital period
ratios of the planet pairs.

\subsubsection{R3: Two planet pairs both in 3:2 MMR}

In R3, we assume the parameters are as follows: the masses of the
three planets are all equal to 5 $M_\oplus$, and the stellar
accretion rate and stellar magnetic field share identical values
with the values considered in R1. However, the speed of the type I
migration is 0.3, ten times the speed considered in R1.

\begin{figure*}
\begin{center}
  \epsscale{1}\plotone{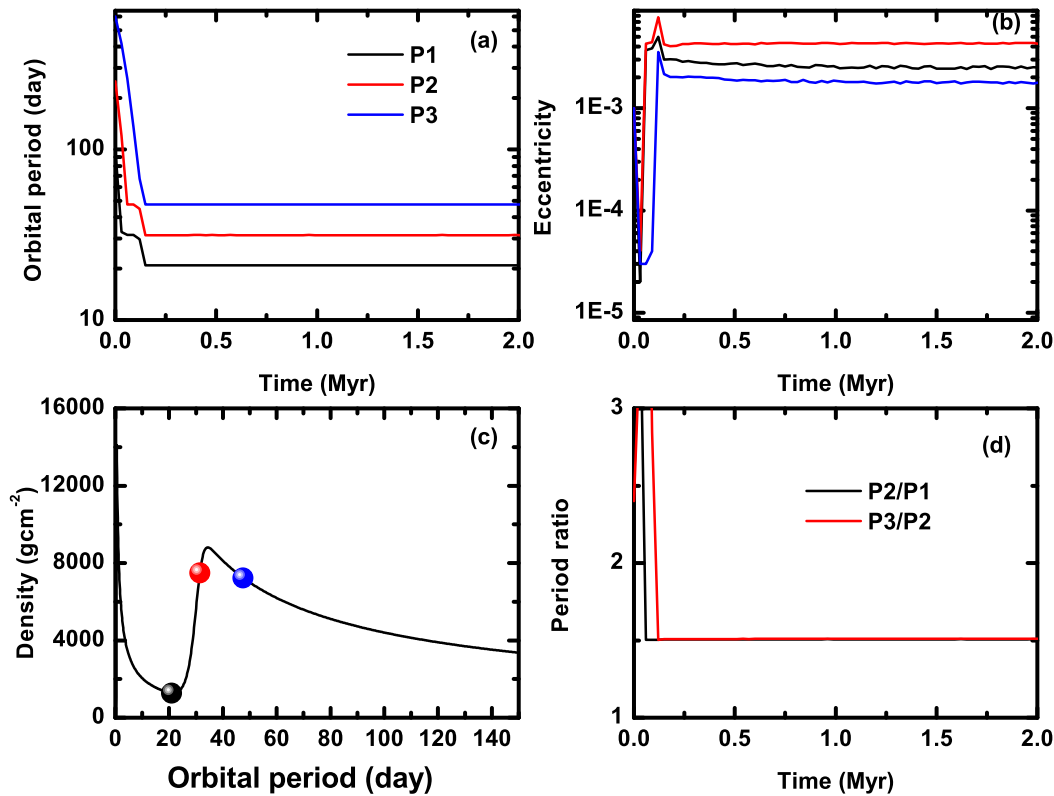}
 \caption{Results for R3. Panels (a), (b), and (d) show the
 evolution of orbital period, eccentricity, and period ratio of
 planets pair, respectively. Panel (c) shows the final configuration
 of three planets. P1, P2 and P3 stand for the innermost, the intermediate and
 outermost planet, respectively. The black solid curve is the
 density profile of the gas disk initially.
 \label{f6}}
 \end{center}
\end{figure*}

In this case, in 0.65\% of all of the simulations, both of the
two-planet pairs are captured in 3:2 MMRs. Figure \ref{f6} shows
that the inner pair falls into a 3:2 MMR quickly. Subsequently, when
P3 arrives, the inner and outer pairs both evolve into 3:2 MMRs. We
note that the innermost planet travels across the region of maximum
density, whereas the other two planets remain in the region where
the gas density remains high. However, we note that the planets'
eccentricities are excited not far from zero, deceasing to
0.002-0.008 during their evolution. In summary, we can conclude that
rapid type I migration drives the three planets toward 3:2 MMRs,
with the orbital periods of the planets being 20.9, 31.5 and 47.5
days. In this case, no comparable planetary analog of the Kepler candidates
is found.

\subsubsection{R4: The inner pair is in a 2:1 MMR, whereas the outer pair is in a 3:2 MMR}

In the numerical simulations performed for R4, we consider the
masses of the three planets to be 5, 5 and 7.5 $M_\oplus$, and the
stellar accretion rate and reduction factor of the type I migration
are equal to the values considered in R2. The stellar magnetic field
is 1.5 KG, higher than the value considered in R2, resulting in a
larger inner hole in the gas disk.

\begin{figure*}
\begin{center}
  \epsscale{1}\plotone{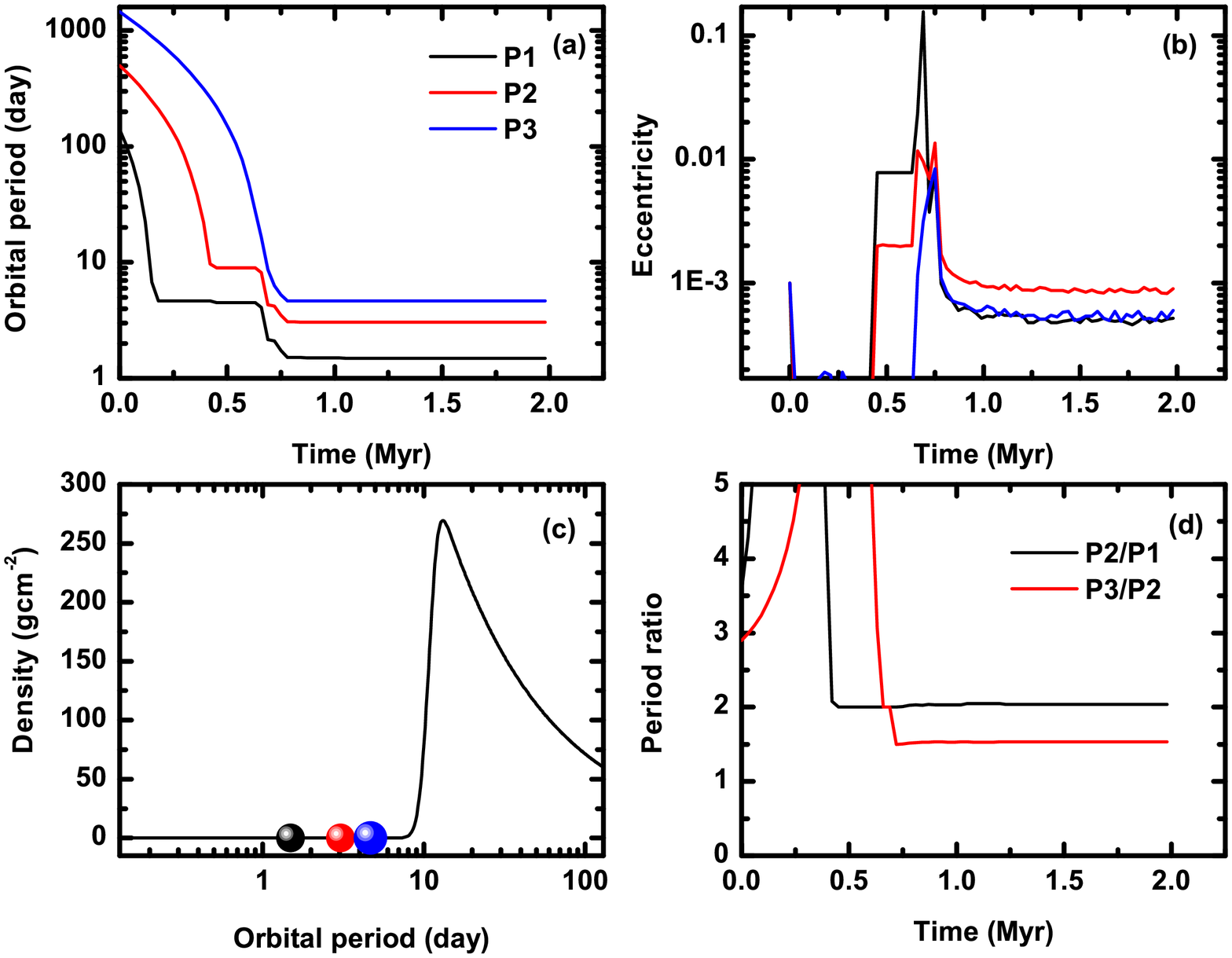}
 \caption{Results for  R4. Panels (a), (b) and (d) show the dynamical
  evolution of orbital period, eccentricity and period ratio of the
  planet pairs, respectively. Panel (c) shows the final configuration
  of three planets. Herein P1, P2 and P3 denotes the innermost, the intermediate
  and the outermost planet, respectively. The black solid curve is the
 density profile of the gas disk initially.
 \label{f7}}
 \end{center}
\end{figure*}

In this group, we notice that 0.32\% of all simulations consist of a
2:1 MMR for the inner pair and a 3:2 MMR for the outer pair.
Therefore, the three companions are locked in the chain of a 3:2:1
MMR. Figure \ref{f7} shows one of the typical simulations. Figure
\ref{f7} shows that the inner pair first falls into a 2:1 MMR at
$\sim$ 0.4 Myr. When the planets are trapped in a 2:1 MMR, their
eccentricities are increased to $\sim$ 0.1 but then are dissipated
by the gaseous disk within 1 Myr. In addition, we find that the
occurrence of P3 indeed triggers the capture of the two outer
planets in a 3:2 MMR, thereby leading to a 3:2:1 MMR for the three
companions. During their evolution, the three planets move across
the region of maximum density and reach the inner hole of the
system, with final orbital periods of 1.5, 3.0 and 4.7 days. This
configuration bears a resemblance to the KOI-1835 system, with three
planets exhibiting orbital periods of 2.2, 4.6 and 6.8 days.

\subsection{Case 2}

Based on the results obtained for G1-G3, we note that the proportion
of planet pairs near 2:1 MMRs is greater than that of planets near
3:2 MMRs when we compare the histogram of the numerical results with
the observations. Thus, a question arises: Is there any likelihood
that the occurrence of additional planets may contribute to the
dynamical evolution of planet pairs in the existing systems over
secular timescales, which could explain the difference between the
simulations and observations? To address this question, we perform
further numerical simulations, initially placing one or more
additional planets into the previous systems that could be trapped
in 2:1 and 3:2 resonances during their evolution. Based on the
results obtained for Case 1, the systems involved in 2:1 and 3:2
MMRs are ultimately chosen for further investigation (see Table
\ref{tb2}), for which it is assumed that the companions settle into
either the closest or the most distant orbits from the central star.
In this case, we perform a total of 24 runs in the simulation,
with the systems divided into the two groups: A1 and A2.

A1: For this group, we investigate how additional planets may affect
the planets captured in or near 2:1 MMRs during their evolution. For
the initialization of the numerical simulations, three subgroups are
considered: (1) a planet with a mass of 2 or 5 $M_\oplus$ is assumed
to be inside the orbit of the innermost planet; (2) a planet with a
mass of 5 or 20 $M_\oplus$ is located outside the orbit of the
outermost planet; (3) two planets are, respectively, placed inside
the orbit of the innermost planet and outside the orbit of the
outermost planet in the system (Table \ref{tb2}).

A2: For this group, we explore the issue of how additional planets
stir up 3:2 MMR systems. For a detailed study, we also classify the
simulations into three subgroups, similar to those case of A1. In
the following section, we briefly summarize the major results.

\subsubsection{The results of A1-A2}
\begin{figure*}[t]
\begin{center}
\includegraphics[scale=0.8,angle=-90]{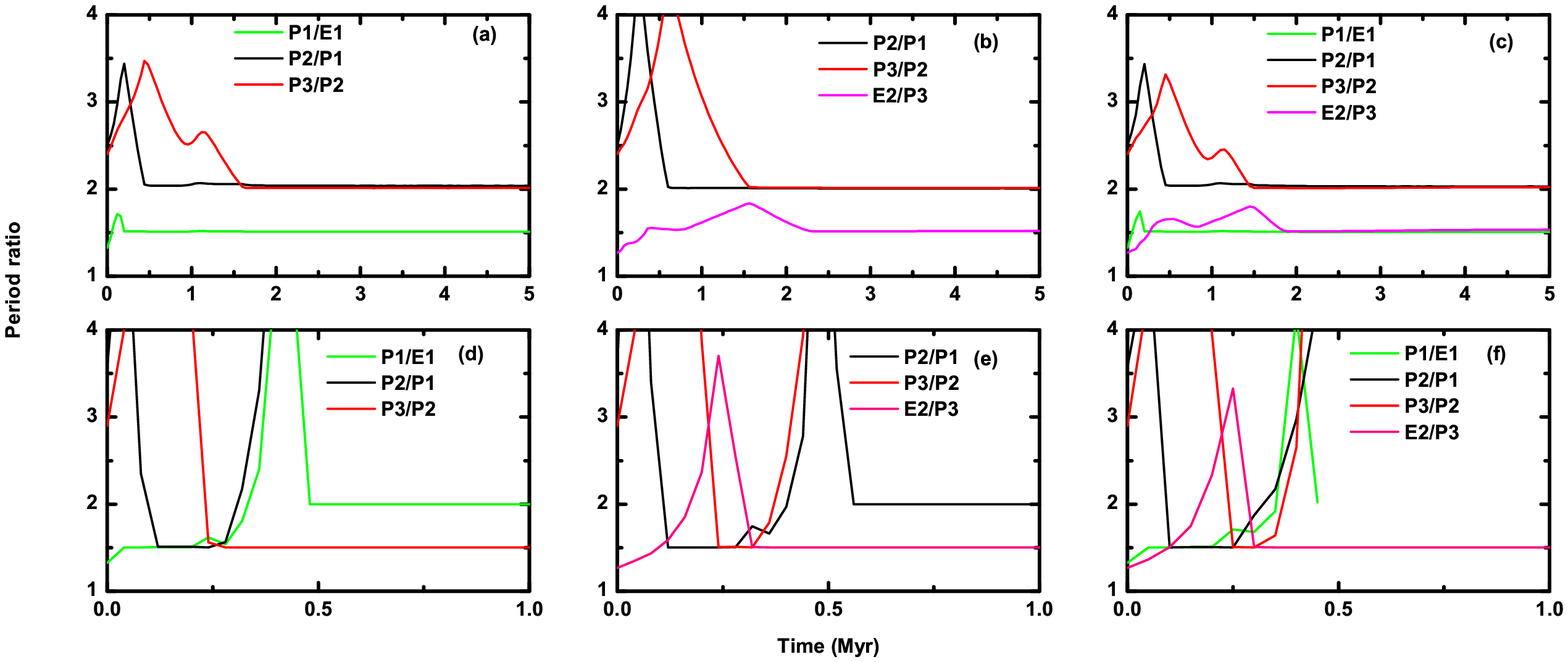}
 \caption{Evolution of period ratios for A1 and A2.
  Panels (a), (b) and (c) show the evolution of period ratio of
  planet pair for A1, respectively. Panels (d), (e) and (f)
  exhibit the cases of A2, respectively.
  Herein E1 represents the supposed planet inhabits inside the innermost planet,
  and E2 denotes the additional planet resides beyond the outermost planet.
  The green, black, red and pink lines show the period ratios of P1/E1
  , P2/P1, P3/P2 and E2/P3, respectively.
 \label{f8}}
 \end{center}
\end{figure*}

Panels (a), (b) and (c) in Figure \ref{f8} show the variations in
the period ratios for a typical run in A1, where E1 represents the
added planet residing inside the orbit of the innermost planet and
E2 denotes the planet inserted outside the orbit of the outermost
planet. As previously mentioned, the adopted three-planet system can
form a configuration approximating a 4:2:1 MMR in Case 1, with a
stellar accretion rate of $2 \times 10^{-8}$ $M_\odot~\rm{yr^{-1}}$,
a stellar magnetic field of 0.5 KG, and a reduction factor of 0.03.
As shown in Figure \ref{f8}, E1 and E2 do not disrupt the 4:2:1 MMR
in the original system, indicating that Laplacian resonance is very
robust. Panels (a) and (b) exhibit orbital period ratios of
P1/E1=1.5 and E2/P3=1.5, respectively, which suggests that E1 or E2
has a good probability of entering into a 3:2 MMR with the nearby
planet (herein P1 or P3, respectively).

Panel (c) shows that E1 and P1 are captured in 3:2 resonance within
a very short timescale, whereas E2 and P3 are locked into 3:2
resonance over a much longer timescale, approximately $\sim$ 2 Myr.
However, it is noteworthy that the resultant configuration of the
3:2:1 MMR is produced for the three planets E2, P3 and P2.
Fortunately, we find that the system KOI-2433, consisting of four
low-mass planets, bears resemblance to the configuration shown in
panel (a), where their orbital periods are 10.0, 15.2, 27.9 and 56.4
days, from innermost to outermost. Again, this finding further
indicates that the formation scenario may be applicable to other
Kepler systems that are close to 3:2:1 MMRs.

In Figure \ref{f8}, panels (d), (e) and (f) show the typical period
ratio in A2. In this run, a three-planet system with a stellar
accretion rate of $2 \times 10^{-8}$ $M_\odot~\rm{yr^{-1}}$, a
stellar magnetic field of 0.5 KG, and a reduction factor of 0.3 in
Case 1 can form a configuration in which both of the planet pairs
are in near-3:2 MMRs. Panel (d) shows that the  E1 and P1 pair is
temporarily trapped in 3:2 resonance; thus, the two planets are
ultimately locked into a 2:1 resonance, whereas P2 and P1
subsequently enter into a 3:2 MMR; however, the resonance is
disrupted within 1 Myr, and the two outer planets P3 and P2 continue
to evolve into a 3:2 resonance.

Panel (e) illustrates that P2 and P1 first undergo a 3:2 resonance
within 0.1 Myr, and then P3 and P2 are locked into this resonance,
which is held for a short period of time. Subsequently, the 3:2 MMRs
for the two pairs (P2, P1) and (P3, P2) are both disrupted owing to
the existence of the additional planet E2. Finally, P2 and P1 enter
into an alternative 2:1 resonance during their dynamical evolution,
whereas E2 and P3 are eventually locked into a 3:2 resonance. Panel
(f) shows that the three pairs of (P1, E1), (P2, P1) and (P3, P2)
are temporarily in 3:2 MMRs over the course of their evolution, but
ultimately, only the outer pair of (E2, P3) is captured in 3:2
resonance.

Based on the above-described results, we conclude that Laplacian
resonance is not easily disrupted by additional planets, and the
additional planets will lead to the formation of 3:2 MMRs; however,
for planetary systems that are originally in two 3:2 MMRs during
their evolution, the 3:2 resonance will disintegrate due to the
interplay of planets.

\section{Conclusions and Discussions}
In this work, we extensively investigated the planetary
configuration formation of systems that are involved in first-order
resonances (e.g., 2:1 and 3:2 MMRs) via numerical simulations. In
total, we performed over 1000 runs, considering systems with
various combinations of planetary mass, stellar accretion rate,
stellar magnetic field, speed of type I migration and additional
planets. Moreover, we also compared our numerical results with the
planetary candidates released by the Kepler mission. We summarize
the major conclusions of our study as follows.

1. Concerning the statistics of the observed data, the proportion of
planet pairs near 2:1 MMRs is $\sim$ 18.0\% for three-planet
systems. However, our numerical simulations show that the proportion
of planet pairs in 2:1 MMRs remains 36.4\% for G1-G3 and 26.0\% for
G4. Apparently, near 2:1 MMR configuration can be formed easily
\citep{lee02} through our formation scenario and it is able to yield
a proportion of planet pairs captured in 2:1 MMRs that is similar to
the observed proportions and should thus provide an explanation of
the behavior of Kepler candidates involved in 2:1 MMRs.

With respect to 3:2 MMRs, the observation results yielded by the
Kepler mission show that the proportion of planet pairs in this MMR
is $\sim$ 7.0\% for three-planet systems. In contrast, this
proportion can reach up to 14.5\% in our simulations due to the
original packed configuration of the systems. This formation
scenario also sheds light on the generation of 3:2 resonant
configurations in Kepler systems.

From our simulations, we get two peak-trough features at 3:2 and 2:1
MMRs which can be seen from the observation data. In this work, we
focus on the planets with mass smaller than 30 $M_\oplus$. Thus, we
believe that the scenario with migrating planets is effective for
the formation of near MMRs configuration for low mass planets
\citep{LT11}. While for massive planets, with a mass larger than 20
$M_\oplus$, the two peak-trough features can be reproduced using in situ
formation scenario with growing planets \citep{Pet13}.

2. The formation of MMRs does not appear to be sensitive to the
stellar magnetic field.

3. Moreover, we find that $\dot M=2\times 10^{-8}~M_\odot~{\rm
yr^{-1}}$, corresponding to the early stages of star formation, is
conducive to 2:1 MMR formation for two planets, whereas $\dot
M=0.1\times 10^{-8}~M_\odot~{\rm yr^{-1}}$, which is indicative of
the late stages of star formation, contributes greatly to 3:2 MMR
formation.

4. Our simulations also suggest that $f_1\geq 0.1$ is a proper value
for the speed of type I migration that may result in the formation
of near-2:1 MMRs, which is consistent with our previously reported
results \citep{Wang12}, whereas $f_1\geq 0.3$ appears to favor the
production of planet pairs in near-3:2 MMRs. To summarize, a slower
speed of type I migration (as low as one-tenth the theoretical
value) plays a vital role in the resonance formation of systems.

The 1:2:4 and 1:2:3 MMRs usually formed when the first planet is
trapped in the edge of the holes in the gas disk. The formation
scenario is similar to that mentioned in \cite{PN08}, the MMR formed
when the planet is captured in the gap of another planet.

Furthermore, compared with our previous work, herein, we have not
explored tidal interactions between a central star and its planets
\citep{LW12,BM13}. In the present work, we merely consider planets
with a mass of less than 30 $M_\oplus$ based on a tidal timescale
\citep{ML04,ZL08}, under which the planets' semi-major axes and
eccentricities will gradually decrease but the period ratios will
change little due to tidal effects. In fact, tidal interactions
would trigger the degradation of first-order resonance systems to
near-resonance systems \citep{lee13}.

In summary, we conclude that the near-MMR configurations such as
those exhibited by Kepler candidates could be produced under this
formation scenario. In addition, our investigation may reveal that
this formation scenario is not only applicable to Kepler systems but
is also suitable for other planetary systems, especially for systems
composed of several short-period planets with low masses. In this
respect, our work should shed new light on the planetary formation
of such systems in general.

\acknowledgments{We thank the referee for the
constructive comments that helped to improve the original
content of this manuscript. W.S. and J.J.H. are supported by National Natural
Science Foundation of China (Grants No. 11273068, 11203087,
11473073), the Strategic Priority Research Program-The
Emergence of Cosmological Structures of the Chinese Academy of
Sciences (Grant No. XDB09000000), the innovative and interdisciplinary
program by CAS (Grant No. KJZD-EW-Z001), the Natural Science
Foundation of Jiangsu Province (Grant No. BK20141509), and the Foundation
of Minor Planets of Purple Mountain Observatory.}

\end{document}